# Adapting Node-Place Model to Predict and Monitor COVID-19 Footprints and Transmission Risks

**Jiali Zhou**
Department of Urban Planning and Design, University of Hong Kong
Address: 8/F, Knowles Building, The University of Hong Kong, Pokfulam Road, Hong Kong
Email: jlzhou@hku.hk
ORCID: https://orcid.org/0000-0002-4056-8595

**Mingzhi Zhou**
Department of Urban Planning and Design, University of Hong Kong
Address: 8/F, Knowles Building, The University of Hong Kong, Pokfulam Road, Hong Kong
Email: mingzhi@connect.hku.hk
ORCID: https://orcid.org/0000-0001-7472-9329

**Jiangping Zhou**
Department of Urban Planning and Design, University of Hong Kong
Address: 8/F, Knowles Building, The University of Hong Kong, Pokfulam Road, Hong Kong
Email: zhoujp@hku.hk
ORCID: https://orcid.org/0000-0002-1623-5002

**Zhan Zhao, Corresponding Author**
Department of Urban Planning and Design, University of Hong Kong
Address: 8/F, Knowles Building, The University of Hong Kong, Pokfulam Road, Hong Kong
Email: zhanzhao@hku.hk
ORCID: https://orcid.org/0000-0001-5170-9608

**Abstract**

The node-place model has been widely used to classify and evaluate transit stations, which sheds light on individuals' travel behaviors and supports urban planning through effectively integrating land use and transportation development. This article adapts this model to investigate whether and how node, place, and mobility would be associated with the transmission risks and presences of the local COVID-19 cases in a city. Similar studies on the model and its relevance to COVID-19, according to our knowledge, have not been undertaken before. Moreover, the unique metric drawn from detailed visit history of the infected, i.e., the COVID-19 footprints, is proposed and exploited. This study then empirically uses the adapted model to examine the station-level factors affecting the local COVID-19 footprints. The model accounts for traditional measures of the node and place as well as actual human mobility patterns associated with the node and place. It finds that stations with high node, place, and human mobility indices normally have more COVID-19 footprints in proximity. A multivariate regression is fitted to see whether and to what degree different indices and indicators can predict the COVID-19 footprints. The results indicate that many of the place, node, and human mobility indicators significantly impact the concentration of COVID-19 footprints. These are useful for policy-makers to predict and monitor hotspots for COVID-19 and other pandemics' transmission.

1. **Introduction**

COVID-19, as a global infectious disease, has become an international concern and unprecedently hit cities worldwide. Since declared a pandemic by the World Health Organization in March 2020, the disease is still continuing to spread in many parts of the world. Hong Kong, for instance, has suffered five surges of locally confirmed COVID-19 caseloads since the first case emerged in the city (Gov, 2022). After initial fear of the unknown health crisis, however, people had gradually considered fighting against or coexisting with the pandemic as a long-lasting task or "new normal". This could be indicated by adjustment in human's perceptions and behaviors, e.g., normality of telecommuting and resuming of some physical social activities rather than universal lockdowns (Awad-Núñez et al., 2021; Molloy et al., 2021). Hence, the underlying mechanism of how the COVID-19 virus spreads out across space and people's response to the interventions could vary in existing time compared to the earlier pandemic periods. There are emerging calls for more targeted and effective countermeasures to make a balance between mitigating infection risks and maintaining essential human activities and well-being.

Existing studies explored how urban physical environment (especially land use patterns and built environment) are associated with the caseloads and potential virus spreading (Li et al., 2021a; Xu et al., 2022). Distribution and density of certain facilities like restaurants, hotels, and bars were typically found to positively impact the local case increase (Benzell et al., 2020; Ma et al., 2022). Other physical features like street connectivity and land use mix are also directly or indirectly correlated to closeness and intensity of humans' physical contacts across space, which potentially contributes to virus transmission (Kan et al., 2021; Nguyen et al., 2020). Against this backdrop, people have started advocating sustainable urban planning and design to prepare for future virus attack or other crises (Megahed and Ghoneim, 2020). Nonetheless, most of studies have addressed the abovementioned linkage based on the citywide or regional-level pandemic progress. Little discussion has been given on more spatially granular footprints of individuals infected and interactions among people in different places.

Considering human movement and their activities could impact social interactions in place, human mobility is another essential topic for understanding the COVID-19 virus transmission -- in existing scholarship, we know that human mobility patterns can well predict the susceptible-infected-recovered process in cities (Chang et al., 2021). On the basis, public transportation systems, where mobility mostly occurred, also offer fitting indicators explaining the pandemic process (Afrin et al., 2021; Mo et al., 2021). In addition, heterogeneity in human movement and virus transmission can be attributed to socio-economic disparities. Population with low income and lower education level, for instance, could be more vulnerable during the pandemic (Chang et al., 2021; Grekousis et al., 2021). However, little research has integrated concerns on land use patterns, built environment, sociodemographic, transportation systems as well as individuals' mobility and interactions into a certain analytical framework to predict the COVID-19 footprints at a local scale.

Learning that the node-place model has been widely used to describe how characteristics of station-served communities (i.e., services and physical settings) could be related to their attractiveness to human activities (Cao et al., 2020; Kamruzzaman et al., 2014), we framed this model upon the context of the COVID-19 pandemic. We involved "mobility" as an extended aspect of the normal node-place model to potentially offer knowledge on how human mobility and activities, urban physical settings, and potential health risks are interacted with one another in the city. Accordingly, this article is expected to achieve at least the three following objectives or contributions:

(1) An extended node-place-mobility (NPM) model is proposed to measure relative importance of stations with regards to the mobility network. This allows us to better consider the complex links between the built environment, land use, and human mobility network characteristics of stations.

(2) We formulate a new metric from detailed visit history of the infected, that is, the COVID-19 footprint. Compared to previous studies that lack detailed epidemiological investigation, the new metric that records individual footprints can help better capture the transmission process and investigate transmission regularity of the COVID-19 virus.

(3) Through the proposed node-place-mobility model, we examine how local station characteristics are correlated with COVID-19 footprints. Specifically, using regression analysis, we show how mobility patterns of people, as measured in mobility network centrality, can help us predict spatial distribution of COVID-19 footprints. This could examine the effectiveness of anti-pandemic interventions and indicate potential transmission risks among people in different places.

The remainders of the paper are organized as follows. The next section (Section 2) is a literature review of the relationship between land use, built environment, human mobility and COVID-19 situation, and the node-place model and relevant theory. Sections 3 and 4 elaborate on the methodology used and an empirical study of Hong Kong SAR, China. Section 5 presents empirical results. Section 6 concludes.

2. **Literature review**

In existing studies, various indicators extracted from different data and methods have been used to describe the physical environment, transportation characteristics, and human mobility in city. Some of them have been found to be predictive of the COVID-19 pandemic progress. We conducted a literature review on relative studies to inspire probable indicators for our measurement. Against this backdrop, the concept of the node-place model was also introduced. Inspired by existing applications of this model, we aim to propose its future extension to involve dynamic indicators for the pandemic prediction.

*2.1 Impacts of land use and transportation characteristics on COVID-19 situation*

There have been many studies discussing how urban physical settings, characterized by land use patterns and built environment features, impact the spreading of COVID-19 virus (Megahed and Ghoneim, 2020; Xu et al., 2022). Types of facilities and functions of locations can generate various transmission risks. For instance, when examining the effectiveness of location-related anti-pandemic countermeasures, school closure and restrictions on social gathering in public spaces may play a role in containing citywide virus spreading (Litvinova et al., 2019). Notably, modeling virus transmission risks across points of interest (POIs) based on mobility data, Chang et al. (2021) found that full-service restaurants and hotels suffered most. Benzell et al. (2020) examined and ranked the relative transmission reduction benefits and social cost of different categories of locations based on smartphone data and consumer preference surveys in the US. Ma et al. (2022) found the density of supermarket and hotel, business land proportion, and park density particularly play a role in explaining different phases of COVID-19 case increase in Singapore. In addition, various land use or built environment characteristics were used to explain the spatial patterns of COVID-19 cases in cities. Kan et al. (2021) examined features including covering nodal accessibility, building density, average building height, green spaces, sky view, and land use. Nguyen et al. (2020) leveraged Google Street View to detect street features and found indicators of mixed land use (non-single-family home), walkability (sidewalks), and physical disorder (dilapidated buildings and visible wires) were associated with larger quantity of COVID-19 cases.

Human mobility and its performances in transport systems have also been found to explain the susceptible-infected-recovered process in cities (Chang et al., 2021; Schlosser et al., 2020). Hence, some indicators describing characteristics of transportation systems and peoples' travel behaviors were fitted for prediction. Manzira et al. (2022) examined how different modes of transport, including traffic volume, bus passengers, pedestrians and cyclists, were associated with the reported COVID-19 infections. In particular, public transport has been considered as a potential high-risk environment for virus transmission due to passengers being confined and interacting in limited spaces and difficulty to detect potentially transmission routes among people (Gartland et al., 2022). Mo et al. (2021) found that partial closure of bus routes, altering departure times, or limiting capacity of buses might help decelerating the virus spread. Malik et al. (2022) examined that subway transport data contributes to the quality of forecasting COVID-19 spread in New York City. In addition, physical features of transport systems would also exert an influence on disease transmission. Afrin et al. (2021) suggested that transit neighborhood development, e.g., flexible bike- and walk-friendly pathways in public spaces could make uncrowded public transit with safe distancing. Li et al. (2021b) found that POI density around railway stations and travel time by public transport to activity centers were connected with the COVID-19 spread.

Overall, though various indicators extracted from land use patterns, built environment, and mobility have been examined, most of the existing discussions were conducted from a citywide or regional scale. Little research has systematically involved them to explore the impact COVID-19 transmission at the local scale based on footprints of the individual COVID-19 infected.

## 2.2 Human mobility and COVID-19

The mobility network has been widely used in air-borne disease studies for detailed transmission modeling and contact-tracing, which examined how human mobility plays an essential role for pandemic prediction. For instance, Mo et al. (2021) proposed a time-varying weighted encounter network to model pandemic spreading in public transportation systems using smart card data and concluded that isolating influential passengers at an early stage can reduce the spreading. Yabe et al. (2020) utilized a contact network based on human mobility trajectories (GPS traces) and web search queries to predict COVID-19 hotspot locations. They also proposed a high-risk social contact index to capture contact density and COVID-19 contractions risk levels. These studies, however, often employed unipartite networks of people and ignored the interactions between people and the locations where human activities take place.

Bhattacharya et al. (2021a) constructed a homogeneous network of locations based on aggregate mobiloty flows and employed the PageRank algorithm (Brin and Page, 1998) to identify the high-risk locations for COVID-19 transmission, but did not capture the individual-level human movements. Considering that the transmission processes take place through contact networks of infected individuals, it is important to consider both individual mobility traces across locations and colocation patterns across individuals. This can be done by constructing and analyzing a heterogeneous bipartite network that connects people and their visited locations using large-scale human mobility data. For example, Zhou et al. (2022) constructed a bipartite people-location network with metro smart card data and calculated the Personalized PageRank scores to identify high risk users and locations regarding COVID-19 transmission.

Taking into account the neighborhood's points of interests (POIs), Chang et al. (2021) introduced a metapopulation susceptible–exposed–infectious–removed (SEIR) model that integrates fine-grained, dynamic mobility networks, using mobile phone data, to simulate the spread of SARS-CoV-2 in ten of the largest US metropolitan areas. They found that a small minority of 'super-spreader' points of interest account for a large majority of the infections. However, as mentioned before, besides land-use, built-environment, human mobility network features, transportation systems also serve as a disseminator of infectious diseases.

In general, land-use, built-environment, sociodemographic, transportation accessibility features, as well as actual human mobility patterns are all needed to capture the dynamics of COVID-19 transmission process accurately. So far, little research has integrated all these aspects well into an appropriate framework for pandemic prediction.

*2.3 Node-place model*

The node-place model has been widely used to describe stations' realization and balances of good services on the network (the 'nodes') with a location that promotes human interactions (the 'place') (Bertolini, 1999). It offers a framework to understand and explore the development potential of station-served areas in cities to improve transit-oriented development (TOD). So far, a variety of node-place measures have been used to capture the features of stations, typically including built environment characteristics (Kamruzzaman et al., 2014) and socioeconomic factors (Zemp et al., 2011). To extend the scope of model, some studies add particular aspects to the node-place model, such as orientation related to street connectivity (Lyu et al., 2016), walkability (Jeffrey et al., 2019), travel network (Dou et al., 2021), accessibility (Cummings and Mahmassani, 2022), design (Zhang et al., 2019), and ridership (Cao et al., 2020).

The node-place model has always been used to classify and evaluate transit stations, which provides insights to learn individuals' travel behaviors and support urban planning through effectively integrating physical environment, sociodemographic, and transportation development. However, use of the model for COVID-19 studies is still lacking. This article will explore how node, place, and mobility would be associated with the transmission risks and presence of the local COVID-19 footprints in the city.

## 2. Methodology

The overall research framework is presented in Figure 1. Specifically, we first collect land-use, sociodemographic, and human mobility data as inputs. Using human mobility data, such as smart card data or GPS data, we construct a bipartite people-location network and calculate the PageRank centrality scores for different stations. Then we introduce the node-place-mobility model, including the index selection and k-means station classification processes. After classifying stations into different clusters using their node, place, and mobility indices, we explore the relationship of these indices and their belonging features with COVID-19 infectors' footprints, taking Hong Kong SAR in China as our case study. The linear regression model is employed to further explore and quantify the relationship.

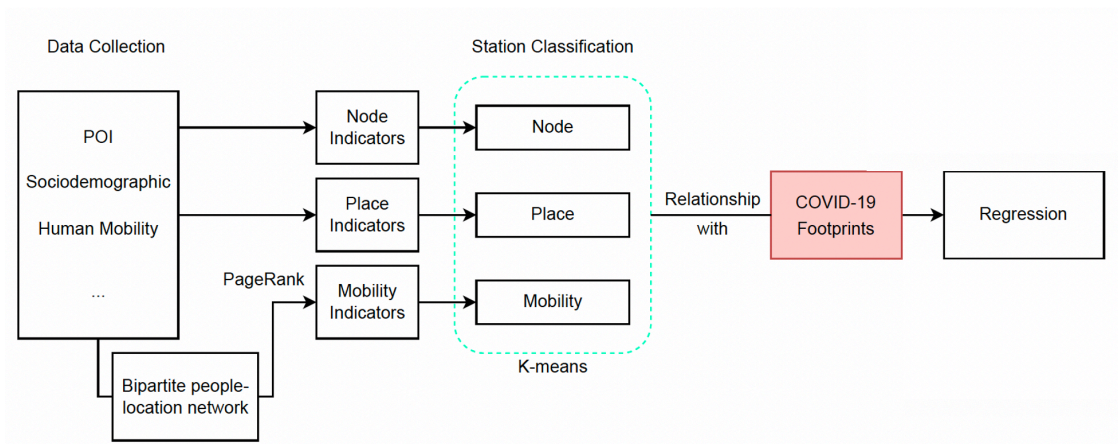

**Figure 1. Research framework**

*3.1 Node-place-mobility model framework*

As mentioned before, the original node-place model ignores the importance of the actual human mobility patterns, that is, how people utilize different locales as a node and a place over time. Only built environment features are considered in the original node-place model. Hence, an extended node-place-mobility model, as shown in Figure 2, is proposed. The node dimension represents the transportation service around the station areas, while place dimension represents the land use features. The new dimension, mobility dimension measures the importance of stations in a bipartite people-location network.

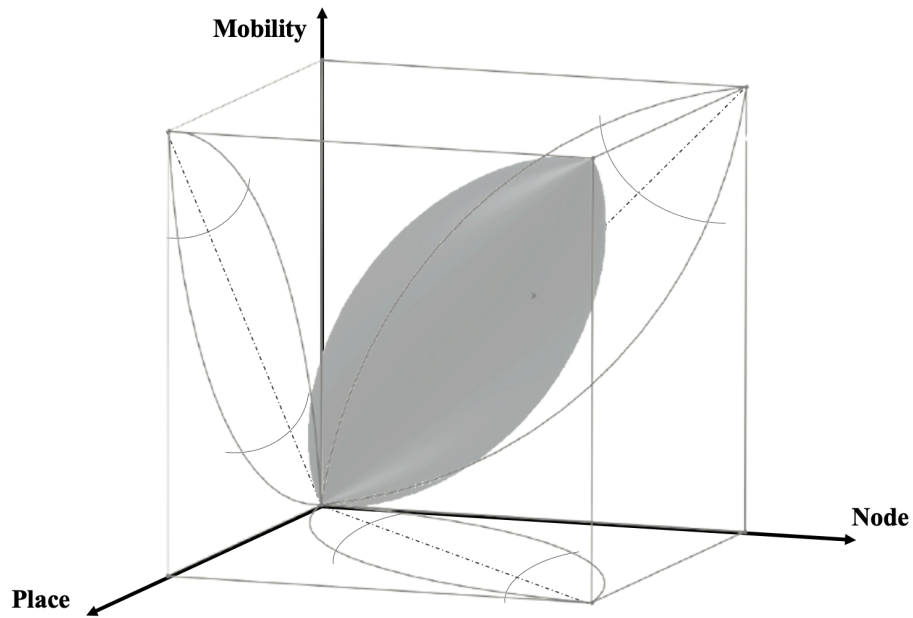

**Figure 2. Illustration of the node-place-mobility network**

Regarding people's first/-last-mile trips around subway stations, the maximum walking distance for pedestrians is found to be around 1.5 km from a subway stations in (Biba et al., 2010; Park et al., 2021). Therefore, the node and place indicators reflecting the land use and transportation accessibility within 1.5 km from the subway stations are selected.

(1) Node indicators

Depending on the form of transportation, node indicators can be classified into two groups: public transit and road networks (Table 1).

**Table 1. Indicators of the node dimension.**

| Node Index | Description |
| --- | --- |
| Number of bus stations | $n_1$ = Number of bus stations within 1.5 km |
| Number of intersections | $n_2$ = Number of intersections within 1.5 km |

(2) Place indicators

The place dimension indicators are shown in table 2.

**Table 2. Indicators of the place dimension**

| Place Index | Description |
|---|---|
| Population density | p1 = population within 1.5 km |
| Percentage of unemployed | p2 = Percentage of unemployed in the population within 1.5 km |
| Percentage of low income | p2 = Percentage of low income (under HK$10000 per month) in the population within 1.5 km |
| Number of POIs | p3 = Number of points of interests (POIs) in the population within 1.5 km |
| Percentage of apartments | p4 = Percentage of apartments in all POIs within 1.5 km |
| Land use mix | p5 = Land use mix |

(3) Mobility indicators

Most existing network-based disease transmission models are based on a homogenous unipartite network of places or persons. A heterogeneous people-location network is constructed in this study to connect people to their visited locations. This is owing to the nature of disease transmission, in which the infected carry viruses to a location and the viruses can survive (on a surface) for certain time. The SARS-COV-2 virus (the virus that causes COVID-19), for example, can survive on a surface for days, according to van Doremalen et al. (2020).

To capture the relationship between two subject types, such as people and locations, a ubiquitous data structure, the bipartite graph, has been proposed. In the bipartite graph, all vertices are categorized into two sets, within which vertices in the same set cannot be directly connected to each other; but they can be connected via a vertex from another set. Treating people and their visited locations as two sets of nodes, a bipartite network can be constructed. Each link describes an individual's visit to a location. The input can be human mobility data of any format, such as cell phone data and transit card data.

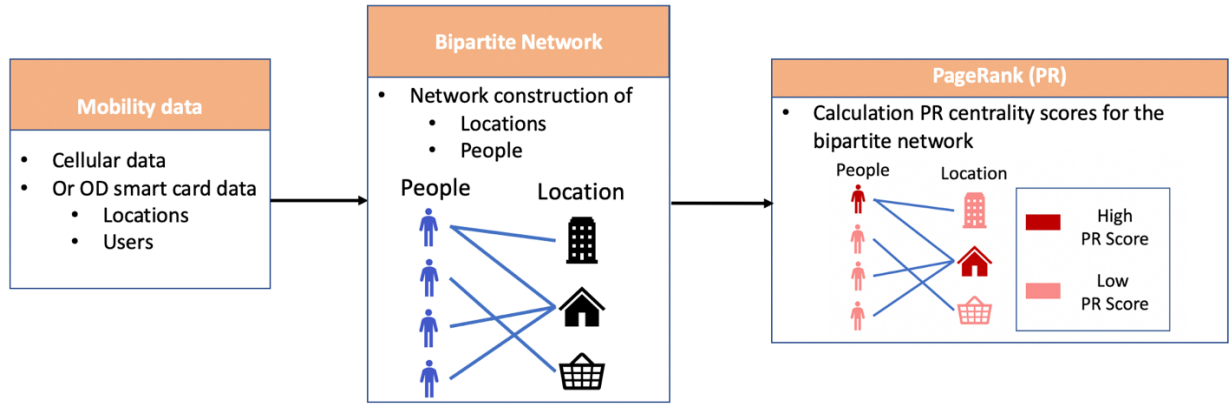

**Figure 3. Construction of bipartite people-location network and calculation of PageRank centrality scores**

As shown in Figure 3, disaggregate-level human mobility data, such as smart card data and cell phone data, is used for bipartite people-location network construction and PR score calculation for transmission risk estimation.

Based on the bipartite network of users and stations, the PageRank centrality scores are calculated for each station. For the calculation, first we assign an equal initial PR score to all nodes. The PR score is transferred between nodes based on the iterations within the network structure of the network via links. The PR score converges after certain iterations. The calculation is conducted by dividing the PR score of the source node by the number of its out-links. The PageRank score of a node $p$ is given as:

$$PR(p) = \frac{1-d}{N} + d \sum_{i=1}^{k} \frac{PR(p_i)}{C(p_i)} \qquad (1)$$

where $N$ is the total number of nodes on the web, $p_i$ is the node that links to node $p$, $d$ is a damping factor, and $C(p_i)$ is the number of out-links of $p_i$. The damping factor $d$ defines the probability of a web user going to an adjacent link, and thus the probability of a surfer skipping to another page is then given by (1-$d$).

The original PR algorithm has been introduced to identify the high-risk locations for COVID-19 transmission (Bhattacharya et al., 2021b), but it is only based on aggregate mobility flows and a homogeneous network of locations. The implementation of the PageRank Algorithm can be described below:

| Algorithm1 PageRank Algorithm |
| --- |
| 1  Input Graph G with N nodes |
| 2  Initialize damping factor d |
| 3  Initialize all nodes in G with original PR score = 1/N |
| 4  While not converged |
| 5     For all node v in the graph, do |
| 6        $PR(p) = \frac{1-d}{N} + d \sum_{i=1}^{k} \frac{PR(p_i)}{C(p_i)}$ |
| 7     End for |
| 8  If the error rate for any vertex in the graph falls below a given threshold, Converged |
| 9  End While |

### 3.2 *Clustering*

After data processing, the stations were clustered into different groups using the k-means clustering methods and node, place, and mobility index values as input. Cluster analysis is used to create classes of Hong Kong's metro stations with the least variance within each group and the most variance between them. Here, K-means cluster analysis is used. The average silhouette approach is used to determine the appropriate number of clusters (Rousseeuw, 1987). The traditional approach would stop here and give management recommendations for the station area. What is not obvious is what role the station plays at the strategic network level. We add a criticality analysis of each station to the cluster analysis to highlight differences between stations within the same clusters and indicate the strategic importance of each station region.

### 3.3 *COVID-19 footprint regression*

In this study, we introduce a new COVID-19 metric, the COVID-19 footprint of the infected, which records all locations the infected visited, according to the Hong Kong SAR government's open data on COVID-19 situations in the city. The reason for introducing this new metric is that the place where people get infected remains unknown due to the lack of detailed epidemiological investigation. Only the places visited by the infected are recorded in the open data. As a result, we assume a uniform distribution across all locations the infectors visited before the positive diagnosis.

### 3.3.1 COVID-19 footprints

Here we have total number of infected cases *n*. We define a person *i,* who has been to $m_i$ distinct locations, the footprint of person *i* at location *j* is:

$$v_{i,j} = \frac{1}{m_i} \tag{2}$$

And the number of footprints from all persons who visited location *j* is:

$$v_j = \sum_{i=0}^{n} v_{i,j} \tag{3}$$

The total number of visited locations by infectors within a specific distance *d* from a station *s* can be expressed as a function *l(d,s),* which is a function of *d* and *s*.

The COVID-19 footprint around that station is defined as the sum of weighted number of visits by the infected to any locations within a certain distance from a subway station, as shown below.

$$COVID-19\ footprints = \sum_{j=0}^{l(d,s)} v_j \tag{4}$$

### 3.3.2 Multi-variate regression models

As mentioned in the introduction and literature review sections, socio-economic factors such as low-income population, unemployed population (Chang et al., 2021; Grekousis et al., 2021), types and density of buildings and land use features (Kan et al. (2021), mixed land use (Nguyen et al. (2020), human mobility and transportation systems (Chang et al., 2021; Schlosser et al., 2020) are found to explain the COVID-19 situations in cities. Based on the previous literature, we conducted multiple linear regression with ordinary least squares to discover individual effects of various factors such as unemployed population, density of apartments, land use mixture, transportation service levels, and human mobility levels on COVID-19 footprints around subway stations. This method allows researchers to investigate the links between a set of indicators and COVID-19 footprints in the context of a larger set of variables. A least squares regression approach was applied.

### 4. Empirical study

### 4.1 Study area

In this paper, we chose Hong Kong SAR, China as the study area. The Mass Transit Railway Corporation (called "MTR" locally) is Hong Kong's primary public transportation system, with the highest frequency and daily ridership across all modes of public transportation in the territory (Legislatice Council Secretariat of Hong Kong SAR, 2016). It carries 41% of the daily passenger trips in Hong Kong. There were 230.9 kilometers of rail track and 98 stations in the system as of February 2022.

Regarding the COVID-19 situation, the Hong Kong Centre for Health Protection gathered daily individual confirmed COVID-19 cases and made them available to the general public online (at https://data.gov.hk). The characteristics of each confirmed case (e.g., age, gender), the type of case (imported, local, close contact with local cases, possibly local, close contact of possibly local cases, and close contact of imported cases), and the buildings or venues where the confirmed cases resided or visited in the 14-day period preceding the day of confirmation are all recorded. The case-related sites are divided into two categories: confirmed cases' dwellings and areas visited by them (visited locations).

Hong Kong was hit badly by the fifth wave of the COVID-19 outbreak, which started from December 26, 2021 in Hong Kong (Hong Kong S.A.R. Government, 2022). After February 5, 2022, the Hong Kong SAR government was unable to track the COVID-19 cases in a timely manner. In this study, we used the data between December 27, 2021 and February 5, 2022, there were 1921 confirmed cases, which covers the beginning and the rising phase of an outbreak, where people's mobility patterns are not impacted significantly due to the mild trend of the transmission.

In this part, we present a case study based on built-environment and real-world large-scale mobility data from Hong Kong MTR around all MTR stations, with the aim to thoroughly explore the relationship between built environment and COVID-19 footprints around subway stations. We have chosen the smart card transaction data from January 21, 2021, which represents the operations of a typical day, without incidents and special events/holidays. We built a network of 98 location nodes and 1.7 million people's nodes. The people node's average degree is 3.6, and location node's is 86163.3.

4.2 Data description

Regarding the exogenous variable, the COVID-19 footprints from infectors, approximately 92% of them are located within 1.5 km from a subway station. The summary of exogenous variable (COVID-19 footprints) and other node, place, and mobility index indicators for the NPM model and the explanatory variables for the linear regression model are included in the table 3 below.

**Table 3. Descriptive statistics of variables**

| Variables | Mean (stdev) |
|---|---|
| COVID-19 footprints | 23.11 (87.67) |
| Population density | 36723 (40340.79) |
| Percentage of unemployed (%) | 0.0009 (0.00038) |
| Number of POIs | 961 (765.43) |
| Percentage of apartments (%) | 85 (7.65) |
| Land use mix | 0.019 (0.019) |
| Number of bus stations | 58 (29) |
| PageRank centrality scores | 0.0068 (0.025) |

4.3 Clustering results

Through k-means clustering, 4 clusters are chosen with the best silhouette value 0.31. At the same time, the relationships among node, place and mobility indexes were also analyzed to reveal the complex links between station-level built environment, human mobility, and COVID-19 footprints around station areas, which are the circled areas within 1.5 km from MTR stations. The four clusters show different trends in the node, place and mobility indexes and also the number of footprints.

In Figure 4, each of the node, place, and mobility indices were compared with COVID-19 footprints around stations. The four different colors represent 4 different clusters, with circle size representing number of footprints around a station, as is shown below. Cluster 1 is the cluster with high node index, low place index, and high mobility index. Cluster 2 is the cluster with high node index, high place index, and high mobility index. Cluster 3 is the cluster with medium node index, medium place index, and medium mobility index. Cluster 4 is the cluster with low node index, low place index, and medium mobility index. The cluster centers in the node, place, and mobility dimension, number of COVID-19 footprints' mean and standard deviation for the four clusters are shown in table 4. The general trends from the results are each index appears to be positively impacting COVID-19 footprints, even when considered with another index.

**Table 4. Summary of four clusters**

| Cluster | Cluster centers | | | Number of COVID-19 Footprints, mean | Number of COVID-19 Footprints, stdev |
|---|---|---|---|---|---|
| | Node | Place | Mobility | | |
| 1 | 0.73 | 0.41 | 0.76 | 81.99 | 108.88 |
| 2 | 0.82 | 0.82 | 0.71 | 69.86 | 106 |
| 3 | 0.57 | 0.69 | 0.55 | 22.35 | 23.37 |
| 4 | 0.19 | 0.36 | 0.43 | 4.66 | 3.57 |

Similarly, Figure 4(a) shows that node, place, and mobility indices have positive impacts on the COVID-19 footprints. Therefore, it appears that the higher any of the node, place, or mobility qualities at a station are, the higher number of COVID-19 footprints are around that station area. However, the figures also show the trend of the node index having an outsized impact on COVID-19 footprints relative to place and mobility. This trend of the node index measures showing the impact from transportation service and accessibility. In Figure 4(b) and c, the typical node-place plot, it is shown that the majority of the MTR stations have balanced node-place-mobility values.

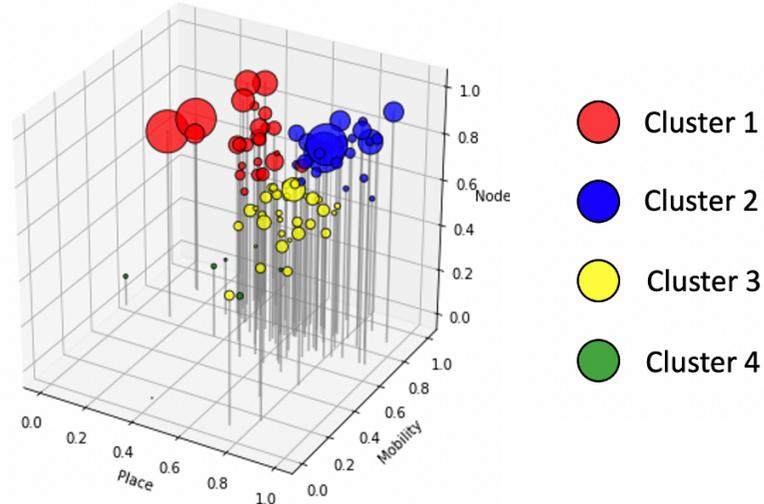

(a) Node-place-mobility indexes for all stations

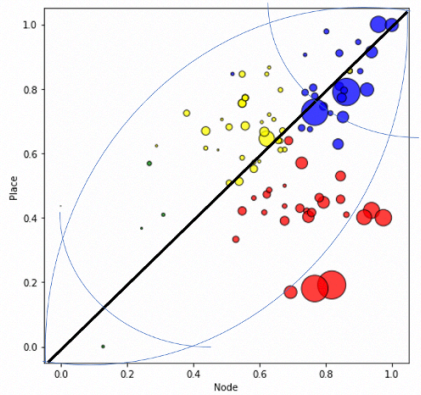

(b) Node-place indexes for all stations

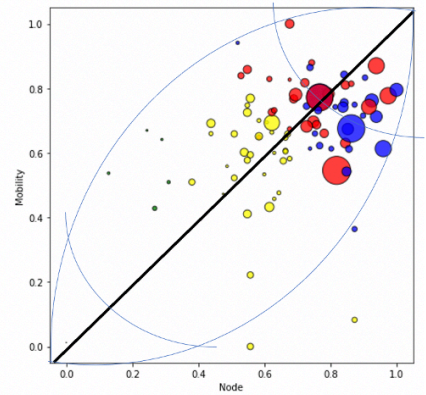

(c) Node-mobility indexes for all stations

**Figure 4. Node, place, and mobility indexes of the stations (with circle size representing number of footprints around the station areas)**

Cluster 1, which is located close to the unbalanced node region, has a higher number of COVID-19 footprints. This means that the areas where the node index value is more dominant than place index value has relatively high number of COVID-19 footprints. One typical example station for cluster 1, Tai Wo Hau, is displayed in figure 5. Tai Wo Hau in Hong Kong is main comprised of public housing apartments, which has lower place index such as land use mix. However, due to its population and the transportation accessibility within 1.5 km, the node and mobility index are still high.

Cluster 2, which is located close to the 'stressed' region (high node and place index values), also has a high number of COVID-19 footprints. The 'stressed' areas, at the top of the line, interaction between transportation and land use dynamics reach the highest value. In fact, in these areas, further development due to restricted space leads to incompatibility in transportation and land use systems. The stressed stations also have a high number of footprints. One example is Central station, which is one of the CBD of Hong Kong. The number of jobs and POIs are high around the Central station, and it has large number of bus connections within 1.5 km. Since it is also one of the job centers in Hong Kong, the mobility index for Central station is also one of the highest.

Cluster 3 has relatively balanced node, place and mobility index. The number of COVID-19 footprints are low for these balanced stations. Tai Wai station, for example, is located right at the outskirt of the traditional Kowloon's CBD area, has reasonable number of bus connections, housing and POI density, as well as medium human mobility level.

Cluster 4 has relatively low node and place index and medium mobility index. There is minimal physical human interaction potential and the intensity and diversity of activities are minimal. As a result, the number of COVID-19 footprints are the lowest for these balanced stations. For instance, Wu Kai Sha is the northeastern terminal station on the Tuen Ma line, and it has relatively low population, POI density, what is more, the transportation accessibility, such as bus stations and road network density there is low, as well as human mobility index.

Another pattern that can be seen in Figure 4 (c) is that the mobility index does not vary a lot for all four clusters, this is due to the small variance of the PageRank centrality scores for the people-location network.

The geographic distribution of stations from different clusters is shown in figure 5. The distribution follows a line-based distribution. For instance, the clusters 1 and 2, two clusters with higher number of COVID-19 footprints, are primarily located in central urban areas with good traffic accessibility and diverse land use in Hong Kong Island and Kownloon's CBD. Cluster 3, which has a medium number of COVID-19 footprints, is mostly located at the edge of the central city area and has a moderate level of development in terms of urban traffic and land use. Cluster 4, with the least number of COVID-19 footprints is located at the edge of the city or close to the end of the rail transit station.

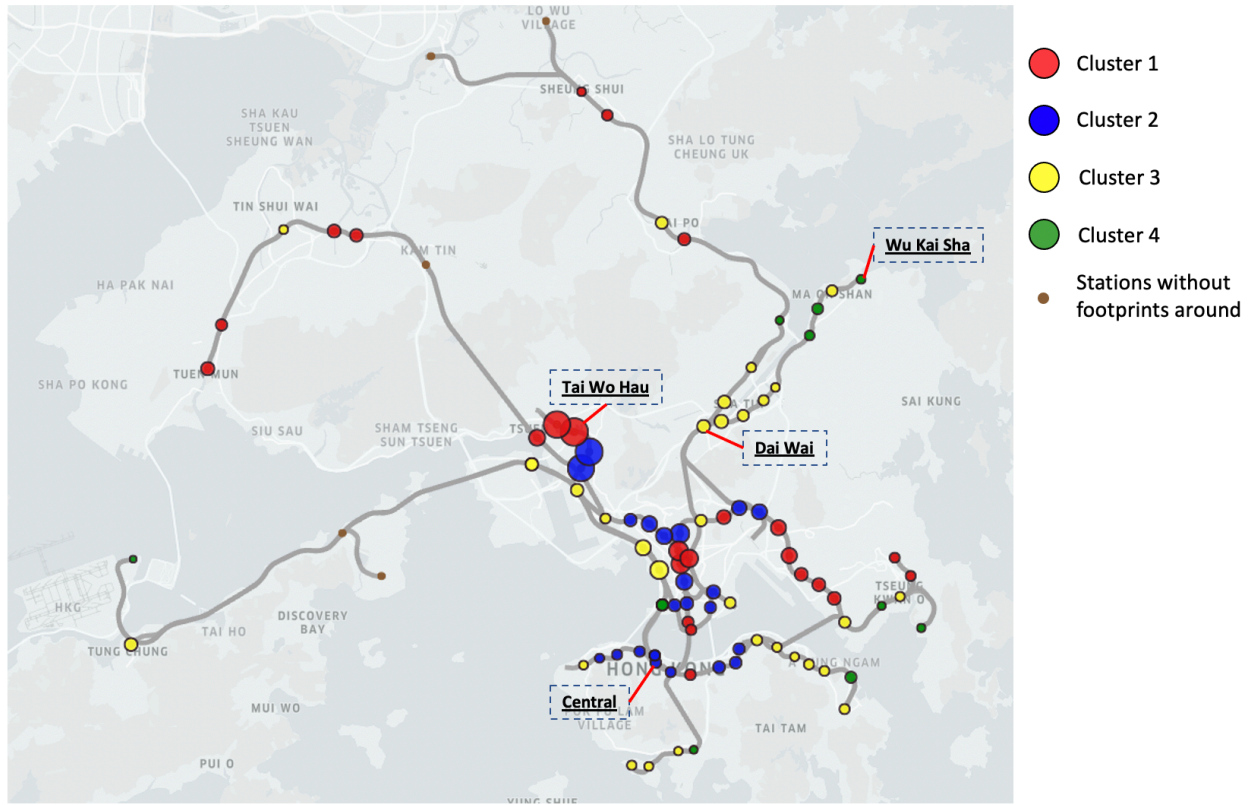

**Figure 5. Spatial distribution of stations from four clusters (with circle size representing number of footprints around the station areas)**

4.4 Linear regression on footprints of COVID-19 infectors

We adopted a two-step regression method as described in Zhou et al. (2021) to fit a series of ordinary least square models using the independent variables mentioned in section 3.3.1, the framework is shown in Figure 6. We first fitted a linear regression model using multiple nodes, place, and mobility indicators. We assume that there exists a linear relationship between COVID-19 footprints and explanatory variables including node, place, and mobility indicators. We also assume higher node index (higher transportation service levels), higher place value (land use and sociodemographic complexity) and human mobility index can lead to more COVID-19 footprints around stations. Specifically, in the final model after the model adjustments and comparison, we included 5 independent variables under the assumptions that unemployed population, more densely distributed apartments, higher land use mixture, higher transportation service levels, and human mobility levels lead to more COVID-19 footprints.

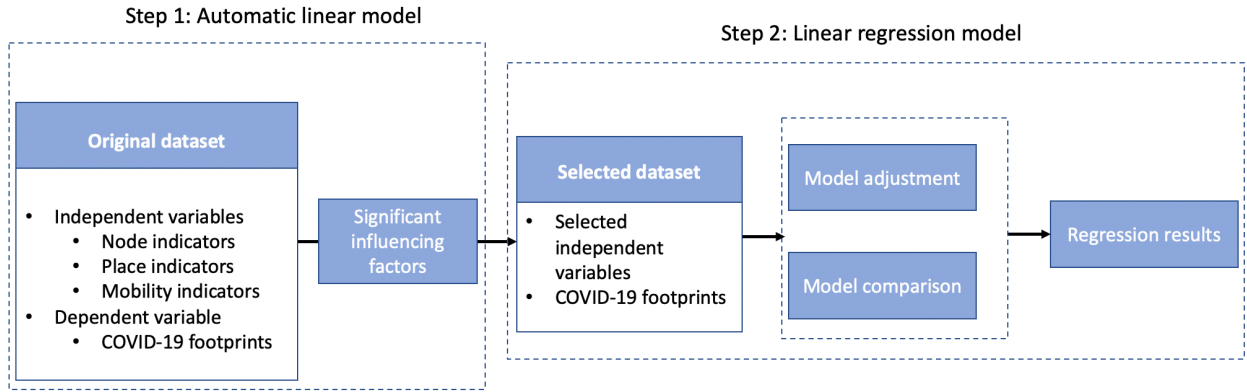

**Figure 6. Regression modeling framework**

The linear regression model on COVID-19 footprints around stations has five independent variables including the percentage of apartments among all POIs within 1.5 km from the subway station, the PageRank score of the subway station in the bipartite people-location network, the land use mix, number of bus stations within 1.5 km from the subway station and the percentage of unemployed in the population within 1.5km.

Because log transformation can better normalize multiple high-variance variables and resulting in a model with enhanced COVID-19 footprints predictability, the log-log form was utilized for the exogenous variable and explanatory variables. For the regression, the logarithm (base 10) of each was calculated.

The model results are shown in table 5. The results show how the node, place, and mobility indicators are associated with COVID-19 footprints around subway stations. Not surprisingly, a larger percentage of housing apartments in all pois within 1.5 km from the MTR stations, more COVID-19 footprints can be observed around stations. The land use mix around a station is positively associated with more COVID-19 footprints. The increase in percentage of unemployed in the general population also indicates more COVID-19 footprints. Regarding the node indicators, the bus station density has positive and significant impacts on the COVID-19 footprints around a station. Last but not least, the higher PageRank centrality scores of the stations in mobility network is, the more COVID-19 footprints could exist.

The results are consistent with the previous findings on the impacts from socio-economic factors (Chang et al., 2021; Grekousis et al., 2021), types of building and density of buildings and land use features (Kan et al. (2021), mixed land use (Nguyen et al. (2020), human mobility and transportation systems (Chang et al., 2021; Schlosser et al., 2020) on the COVID-19 situations in cities. The results of the regression results indicated the positive impact of node, place, and mobility on COVID-19 footprints. In each case high index scores were correlated with more COVID-19 footprints.

**Table 5. Linear regression results estimating COVID-19 footprints.**

| Variable | Coeff | t-value | P-value | Sig. |
|---|---|---|---|---|
| Intercept | 9.299 | 10.965 | < 2e-16 | *** |
| Percentage of apartments | 7.687 | 4.815 | 6.34E-06 | *** |
| Percentage of unemployed | 74010.000 | 2.525 | 0.0134 | * |
| Land use mix | 0.610 | 4.24 | 5.66E-05 | *** |
| Number of bus stations | 0.219 | 2.483 | 0.015 | * |
| PageRank centrality | 0.136 | 1.784 | 0.078 | . |
| | | | | |
| Number of observations | 85.000 | | | |
| Adjusted R-squared | 0.456 | | | |

Note: Asterisks indicate significance: '.' significant level at 0.1 level, '*' significant at 0.01 level, '**' significant at 0.001 level, '***' significant at 0.0001 level

## 5. Discussion and Conclusions

In the existing scholarship, authors have examined how individual local land use and transportation characteristics, and human mobility patterns impact the COVID-19 situation in cities. Previous studies identified important land use features including population density, buildings of specific types, and socio-economical characteristics, transportation features including connections and frequency, and mobility for COVID-19 situation. Nonetheless, most studies have focused on the above-mentioned relationship based on pandemic progress at the city or regional level. Little attention has been paid to more spatially granular footprints of infected individuals and interactions among people in different places, which is critical for more targeted and effective countermeasures to strike a balance between infection risk mitigation and essential human activities and well-being in the post-COVID era. Also, little research has combined land use patterns, the built environment, sociodemographic, transportation networks, and individual mobility and interactions into a specific analytical framework to estimate COVID-19 footprints at a local scale. What is more, few studies have been conducted in the context of transit-reliant cities such as Hong Kong. The node-place model has provided a useful framework for us to conceptualize and quantify different locales' functions provided to these cities and transit services supplied by these cities. In theory, the functions and services should match each other. In this article, we introduce a third dimension to the model: how riders patronized different locales, that is, we have adapted the existing node-model model. We operationalized the adapted model using empirical data from Hong Kong and illustrated that the model can be used to predict and monitor COVID-19 footprints and transmission risks.

Specifically, the node index describes the number of bus stops and intersections around a subway station. Place index indicates the population density, number of unemployed, number of POIs, percentage of apartments among POIs, and land use mix. Mobility index represents the PageRank centrality scores of the stations in a people-location network. Based on the quantification, we categorized different stations into 4 groups and fitted regression models to check how the node, place, and mobility influence COVID-19 footprints and transmission risks. We found that the higher any of the node, place, or mobility qualities at a station are, the higher number of COVID-19 footprints are around that station area. However, the results also show the trend of the node index having an outsized impact on COVID-19 footprints relative to place and mobility. This trend of the node index measures showing the impact from transportation service and accessibility. The regression findings reveal that node, place, and mobility have a beneficial impact on COVID-19 footprints. High index scores are linked to more COVID-19 footprints in each case.

Specifically, to capture the impacts from the aforementioned local land use and transportation characteristics, and human mobility patterns features simultaneously, we proposed a novel node-place-mobility model, combining the static land use and transportation accessibility with the human mobility dynamics in cities for subway station area classification. Through the proposed node-place-mobility model, the analysis uses these indices to identify and classify subway stations to different groups, and employs regression to identify important node-place-mobility variables impacting COVID-19 footprints. The node index characterizing transportation service, the place index representing land use characteristics, and the actual human mobility index, indicating the importance of stations in the bipartite people-location network, showed positive impacts on COVID-19 footprints.

The specific results of the regression models identified a number of characteristics of stations that have significant impact on COVID-19 footprints. They suggest that more COVID-19 footprints are likely to be attained at stations that have (1) more bus stations around, (2) a high PageRank centrality in bipartite people-location network, (3) a higher share of apartments in all POIs around, (4) higher land use mix, and (5) higher share of unemployed population. The results report similar trends mentioned in previous findings regarding COVID-19 situations in cities.

A better developed understanding of what drives COVID-19 footprints at the local level can help us understand the system, including where to allocate testing, tracking, and medical resources for infectious disease containment. Vulnerable neighborhoods in a transit-reliant city, for example, with high density of apartments and unemployed population, can be the focus of government monitoring with regards to COVID-19 risks. The findings contribute to a better understanding of the factors for COVID-19 footprints; the resulting insights could be used to identify COVID-19 and other infectious disease's hotspot and help with COVID-19 monitoring. Our work also provides some insights for future urban planning and design. For instance, the popular destinations reflected by footprints of the infected infer that some stations and their surroundings might offer better facilities, services and/or opportunities than others to attract a larger quantity of riders/people. The findings indicate inequities in the distribution of and accessibility to essential services and facilities in the city.

Some limitations of the study include the current node-place-mobility model is still representing the static situation of the subway station's surroundings and human mobility pattern. The dynamic transmission process of infectious disease and its interactions with land use, transportation and human mobility factors are not thoroughly modelled, especially the day-by-day pattern. Future work could expand on and attempt to adopt the time dimension for exploration and modeling. Also, with more detailed epidemiological of COVID-19 infectors, multiple regression models can be developed for locations of different types, for example, regression models for residential areas' COVID-19 footprints and office areas' COVID-19 footprints can be further explored. Given more detailed mobility data (from GPS or LBS tracking), future research can focus on developing models at the building level, which provides even higher granularity for the identifying COVID-19 footprints' hotspot in cities and helps with more precise and accurate COVID-19 containment and control strategies.

van Doremalen, N., Bushmaker, T., Morris, D.H., Holbrook, M.G., Gamble, A., Williamson, B.N., Tamin, A., Harcourt, J.L., Thornburg, N.J., Gerber, S.I., Lloyd-Smith, J.O., de Wit, E., Munster, V.J. (2020) Aerosol and Surface Stability of SARS-CoV-2 as Compared with SARS-CoV-1. *New England Journal of Medicine* 382, 1564-1567.

Xu, G., Jiang, Y., Wang, S., Qin, K., Ding, J., Liu, Y., Lu, B. (2022) Spatial disparities of self-reported COVID-19 cases and influencing factors in Wuhan, China. *Sustainable Cities and Society* 76, 103485.

Yabe, T., Tsubouchi, K., Fujiwara, N., Wada, T., Sekimoto, Y., Ukkusuri, S.V. (2020) Non-compulsory measures sufficiently reduced human mobility in Tokyo during the COVID-19 epidemic. *Scientific Reports* 10, 18053.

Zemp, S., Stauffacher, M., Lang, D.J., Scholz, R.W. (2011) Classifying railway stations for strategic transport and land use planning: Context matters! *Journal of transport geography* 19, 670-679.

Zhang, Y., Marshall, S., Manley, E. (2019) Network criticality and the node-place-design model: Classifying metro station areas in Greater London. *Journal of transport geography* 79, 102485.

Zhou, J., Wu, J., Ma, H. (2021) Abrupt changes, institutional reactions, and adaptive behaviors: An exploratory study of COVID-19 and related events' impacts on Hong Kong's metro riders. *Applied Geography* 134, 102504.

Zhou, J., Zhao, Z., Zhou, J. (2022) Quantifying COVID-19 transmission risks based on human mobility data: A personalized PageRank approach for efficient contact-tracing University of Hong Kong.
26